\documentclass[pdflatex,sn-mathphys-num,iicol]{sn-jnl}

\usepackage{graphicx}%
\usepackage{multirow}%
\usepackage{amsmath,amssymb,amsfonts}%
\usepackage{amsthm}%
\usepackage{mathrsfs}%
\usepackage[title]{appendix}%
\usepackage{xcolor}%
\usepackage{textcomp}%
\usepackage{manyfoot}%
\usepackage{booktabs}%
\usepackage{algorithm}%
\usepackage{algorithmicx}%
\usepackage{algpseudocode}%
\usepackage{listings}%
\usepackage{caption}
\usepackage{subcaption}

\usepackage{paralist}
\usepackage[shortlabels]{enumitem}
\usepackage{amsmath}
\usepackage[nameinlink]{cleveref}

\theoremstyle{thmstyleone}%
\theoremstyle{thmstyletwo}%

\theoremstyle{thmstylethree}%

\begin{document}

\title[Pairwise Comparison for Bias Identification and Quantification]{Pairwise Comparison for Bias Identification and Quantification}


\author*[1]{\fnm{Fabian} \sur{Haak}}\email{fabian.haak@th-koeln.de}

\author[1]{\fnm{Philipp} \sur{Schaer}}\email{philipp.schaer@th-koeln.de}

\affil*[1]{\orgname{TH Köln}, \orgaddress{\street{Gustav-Heinemann-Ufer 54}, \city{Cologne}, \postcode{50968}, \country{Germany}}}


\abstract{


Linguistic bias in online news and social media is widespread but difficult to measure.
Yet, its identification and quantification remain difficult due to subjectivity, context dependence, and the scarcity of high-quality gold-label datasets. 
We aim to reduce annotation effort by leveraging pairwise comparison for bias annotation. 
To overcome the costliness of the approach, we evaluate more efficient implementations of pairwise comparison-based rating.
We achieve this by investigating the effects of various rating techniques and the parameters of three cost-aware alternatives in a simulation environment. 
Since the approach can in principle be applied to both human and large language model annotation, our work provides a basis for creating high-quality benchmark datasets and for quantifying biases and other subjective linguistic aspects.

The controlled simulations include latent severity distributions, distance-calibrated noise, and synthetic annotator bias to probe robustness and cost-quality trade-offs. 
In applying the approach to human-labeled bias benchmark datasets, we then evaluate the most promising setups and compare them to direct assessment by large language models and unmodified pairwise comparison labels as baselines.
Our findings support the use of pairwise comparison as a practical foundation for quantifying subjective linguistic aspects, enabling reproducible bias analysis.
We contribute an optimization of comparison and matchmaking components, an end-to-end evaluation including simulation and real-data application, and an implementation blueprint for cost-aware large-scale annotation. }




\maketitle
\section{Introduction}
Online media, whether in the form of news articles written by professional journalists or posts published by regular users, is prone to carrying linguistic biases~\cite{saez-trumper_social_2013,wolton_are_2019}. 
Manifesting in non-neutral lexical choice, framing, presupposition, and at times derogatory or defamatory language~\cite{spinde_media_2024}, such bias is consequential in news contexts where it shapes attention, trust, and downstream decisions.
Throughout this paper, we use \emph{bias} to refer to such linguistic bias, i.e., systematic non-neutral wording, framing, or presuppositions that favor or disfavor certain actors or groups.

Bias identification (``\emph{Is the text biased?}'') and quantification (``\emph{To what extent is the text biased?}'') depend on the recipient's background, preferences, and prior beliefs~\cite{spinde_media_2024}.
Due to this, annotator variability and context dependence, bias is subjective and difficult to identify and quantify for both human annotators and automated systems.
Creating high-quality annotated datasets requires costly briefings and training of expert annotators, and having multiple annotators label each text further raises the cost.
This leads to the problem that high-quality gold labels for bias are scarce, hindering the development of automated identification and quantification systems.
Although using a large language model (LLM) for annotation can lower overall costs, synthetic judgments by LLMs are prone to label- or prompt-specific bias and errors. 

While subjectivity hinders the identification of bias as a dichotomous feature, quantifying bias on a continuous scale is even more problematic when annotators lack a common baseline. Quantifying bias using an absolute score is therefore challenging for both LLMs and humans.
A way to compare subjective aspects to help overcome this problem is to elicit \emph{relative judgments}, such as \emph{pairwise comparisons}, and to infer an underlying scale. 
In a pairwise comparison, an annotator (human or model) is presented with two items (texts) and decides which one exhibits more of the target feature (here: bias).
Choosing between two options is considered easier than putting a feature level on a scale.
Aggregating these decisions with a rating system yields a ranking of texts by bias severity.

These positive features of pairwise comparisons motivate our research on pairwise comparisons as a more scalable and interpretable alternative.
Our goal is to evaluate whether pairwise comparison is an effective way to quantify linguistic bias, particularly compared to direct human or LLM assessment.
We further propose and simulate more efficient pairwise comparison approaches.
The main problem with pairwise comparisons is the scaling of the number of comparison operations required.
Ideally, quantification would be based on a complete set of pairwise comparisons, which requires $O(n^2-n)$ relative assessments.
Even with the comparatively lower cost of LLM judgments, the default approach to pairwise comparisons becomes prohibitively expensive.
By increasing the efficiency of pairwise comparisons, we promote a more effective quantification of biases and other subjective linguistic aspects, as well as the use of the method to create high-quality benchmark datasets with human judgments.
While the use of pairwise comparison with human annotation is another promising aspect, our experiments focus on synthetic annotations and the application with LLM judgments for effective, cost-efficient bias detection and quantification. 
Therefore, applying this framework to human annotation is outside the scope of this work.

In this work, we optimize and evaluate the use of LLM-driven pairwise comparison for the identification and quantification of bias.
We provide \begin{inparaenum}[(a)]   
\item an overview of matchmaking, comparison, and rating components of a pairwise comparison framework,
\item an optimization of cost-aware pairwise comparison implementations tested with a controlled simulation setup, and
\item an evaluation of LLM-driven pairwise comparison for bias detection and quantification on expert-annotated datasets.
\end{inparaenum}

\section{Theory of Pairwise Comparison Rating}\label{sec:PC}
The principle of using pairwise comparison to quantify a subjective aspect has a long history in psychometrics and preference learning, leading to the development of a range of comparative rating systems~\cite{thurstone1927law,bradley1952rank,luce1959individual,elo1978rating,glickman1999parameter,herbrich2007trueskill}. 
Direct assignment of a quantitative bias measure to a text requires a rater, whether human or LLM, to compare the text with a hypothetical baseline at varying levels of the scale.
This makes the task difficult and computationally demanding.
By leveraging pairwise comparison, we can break down the problem into smaller, simpler tasks by taking ``two at a time, if you are unable to handle more than that'' \cite{koczkodaj_new_1993}.
Pairwise comparisons are also foundational in information retrieval and human evaluation protocols because they are simpler and yield higher inter-annotator agreement than absolute scoring \cite{joachims2002optimizing}. 
Elo rating systems are increasingly used to compare the outputs of Large Language Models (LLMs) through paired comparisons~\cite{boubdir2023elo}.
With the advent of the \emph{LLM-as-a-judge} paradigm, pairwise schemes have been used to model evaluation, for example, MT-Bench and Chatbot Arena Elo~\cite{zheng2023judging,zheng2023arena}. 
Pairwise comparison has also been used for the quantification of subjective criteria, including readability~\cite{engelmann_arts_2024}, perceived safety~\cite{costa_scoring_2023}, writing quality~\cite{shibata_lces_2025}, perceived violence~\cite{ji_visual_2019}, and, lately, bias in query suggestions~\cite{haak_investigating_2024}.

Pairwise comparisons can be best illustrated by explaining the Elo rating system, which is most commonly used in the chess domain.
The principal idea is that players start with an initial Elo score and compete in matches.
Based on the outcome of these matches, they receive or lose Elo scores according to a probability-based reward-penalty calculation. 
The assumption is that the probability of a stronger player winning against a weaker player is higher than the probability of the weaker player winning.
In an Elo system, a win against a higher-ranked player is rewarded with a higher Elo gain than a win against a lower-ranked player, and vice versa for losses.
Typically, players are paired against those of similar rank. 
As the number of matches increases, a chess player's Elo rating converges to their actual playing strength \cite{boubdir2023elo}.

The basic principles of this can be applied to the task of quantifying subjective aspects, such as bias.
The outcome of a ``match'', a comparison, is determined by a labeling decision made by a human or an LLM.
Similar to how the outcome of a chess match is non-deterministic and depends on many factors related and unrelated to chess, a judgment of which of two texts is more biased depends on the judge, their perception of the texts, their understanding of bias, the task, and many other factors.
However, over a number of pairwise comparisons, the scores of several texts will converge towards a generally agreeable ranking. 

Pairwise comparison consists of three major building blocks: a \emph{rating} system, which is based on data from a \emph{matchmaking} system that determines comparison pairs, and the \emph{comparison component} that determines the outcome of the comparisons.

\textbf{Rating. }
Two general approaches can be identified for pairwise-comparison-based rating, which mainly depend on the available data.
The Elo approach is an \emph{online} or \emph{streaming} rating system, in which scores are updated after every match.
This means that matchmaking can be efficient and match pairs in rounds, where in each round, similarly scored texts are compared.
\emph{Offline} or \emph{batch} rating systems, such as different Bradley-Terry implementations, take a set of pairwise comparisons and fit latent strengths (which quantify bias, in our case) by iterative maximum likelihood estimation~\cite{bradley1952rank, hunter_mm_2004}.
The scores of online rating systems are influenced by the sequence of events. 
In contrast, offline ratings are optimized by utilizing the complete set of comparison data and are unaffected by the sequence in which matches occur.
While offline rating typically yields more accurate scores on the same set of comparisons, online rating has the advantage of better-controlled matchmaking and reduced computational effort.

\textbf{Matchmaking. }
The process of assigning pairs for comparison is called matchmaking.
There is a wide range of approaches to matchmaking, primarily depending on the choice of the rating system and the comparison method. 
For this publication, we evaluate the effectiveness of random matchmaking and matching pairs based on score similarity. 

\textbf{Comparison. }
In pairwise comparison, two items are typically compared qualitatively or quantitatively.
Since we are using pairwise comparison to simplify labeling decisions, we omit quantitative comparisons.
Accordingly, we assess which of a given pair of texts exhibits more bias.
Typically, enhancing the efficiency of the evaluation process involves either acquiring a comprehensive set of pairwise comparisons or securing multiple evaluations from different annotators for each pair of contenders.
In \Cref{sec:optimization}, we provide more details on how we can achieve both objectives while reducing the total cost of comparisons.     

\section{Optimization of Pairwise Comparison}\label{sec:optimization}
If cost and effort are not to be considered, a complete set of pairwise comparisons for \(n\) texts by \(m\) annotators (for $m\times n(n-1)/2$ total comparisons) with an offline rating approach would be the best choice for almost all application scenarios.
However, this comes with a large number of costly comparisons, the majority of which do not significantly increase the quality of the produced ratings.
To address this, we introduce a set of comparison variants that aim to reduce the number of necessary comparisons while maintaining high effectiveness compared to a high-cost baseline.
Our objective is to identify the optimal set of candidates to be evaluated on human-annotated benchmark datasets (\Cref{sec:application}) by simulating various cost-sensitive variants under multiple hyperparameter configurations and dataset conditions.
While simulation does not provide direct proof of effectiveness, we can isolate the implementation setting from less controllable variables, such as annotators' (or LLMs') capabilities to interpret and correctly judge the investigated phenomenon, as well as the dataset's variability and quality. 
Accordingly, the simulations test whether cost-aware variants retain ranking quality under noise and annotator bias.

We consider three cost-aware variants. 
\begin{inparaenum}[(a)] 
\item A \emph{streak pruning} strategy applies early stopping: once an item records $prune_{rounds}$ wins (or losses) against at least a minimum number of $d$ distinct opponents, it is classified as confidently high (or low) and removed from further matchmaking. 
\item The \emph{tail pruning} variant begins pruning after $w$ regular rounds of comparisons.
Then, after each round, a fixed percentage $prune_{perc}$ of participants from both tails is dropped from matchmaking, focusing computation on differentiating the middle band. 
\item A \emph{listwise} variant transforms the pairwise comparison task into a ranking task for small groups of $k$ texts, with optional overlap $ol$ between the groups.
The returned ranking yields $k(k-1)/2$ implied pairwise comparisons. 
One listwise call can thus approximate many pairwise calls, akin to listwise learning-to-rank \cite{cao2007listnet}. 
\end{inparaenum}

\textbf{Simulation Setup. }
In order to assess the strategies within a regulated setting, we use a simulation environment.
To do so, we create datasets with 1,000 entries that have latent scores on a 1-1000 scale, representing a characteristic such as bias severity, under various distribution assumptions.
We simulate linear (approximately uniform), bimodal (two separated modes), and normal (unimodal) distributions.
These latent distributions are chosen to reflect plausible scenarios in bias annotation.
A roughly uniform distribution corresponds to a corpus that spans the full range from unbiased to highly biased texts.
A bimodal distribution captures polarized settings, for example corpora composed of mostly neutral content and a cluster of highly biased items.
A normal distribution models the case in which most texts exhibit low to moderate bias, with relatively few instances at the extremes.
This allows us to assess how implementation variants behave under different assumptions about the underlying distribution of bias severity.

\begin{figure}[t]
    \centering
    \includegraphics[width=1\linewidth]{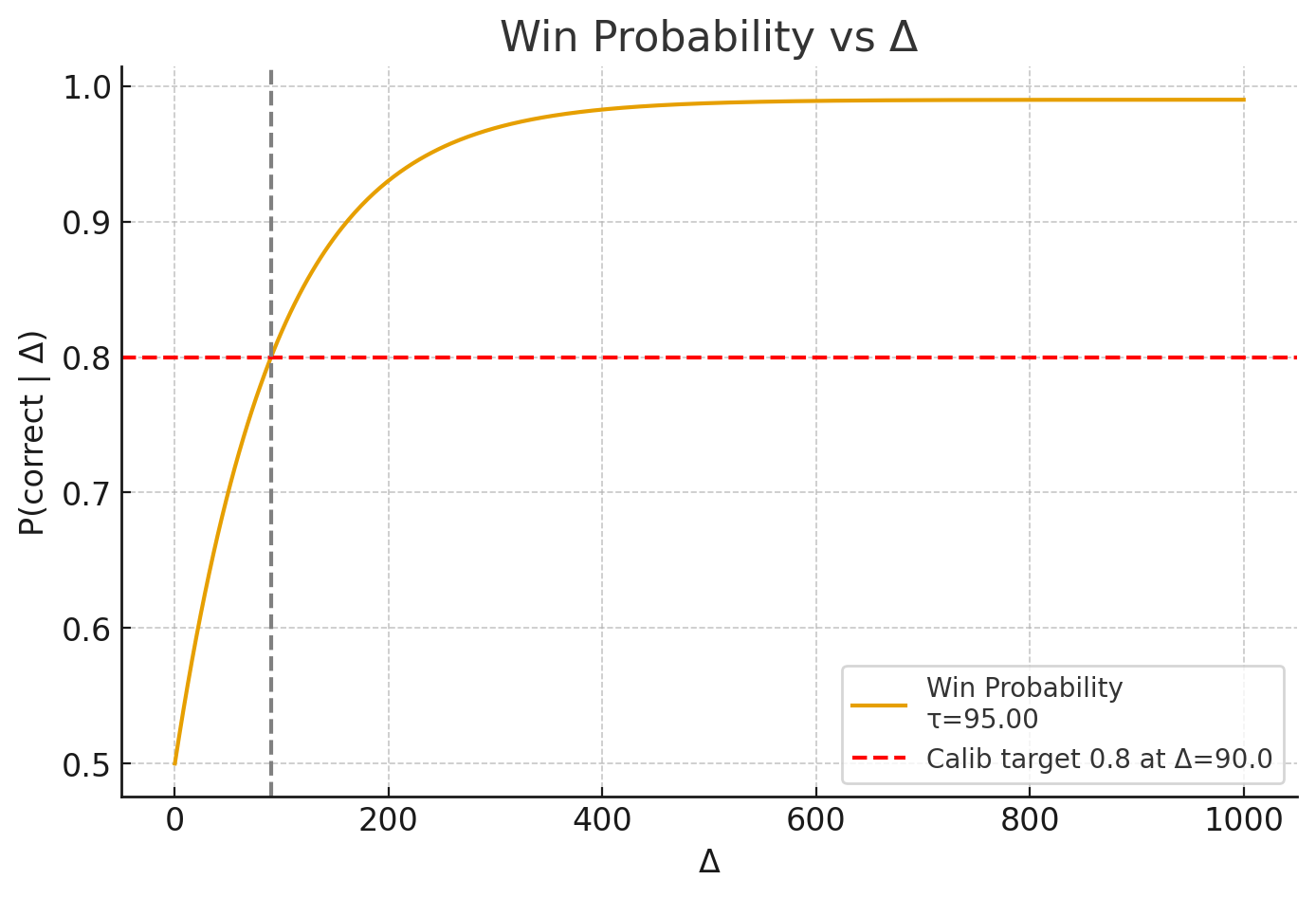}
    \caption{Win probability for a given score $\Delta$ in the simulations' settings.}
    \label{fig:win_prob}
\end{figure}

The outcome of pairwise comparisons is determined by the latent scores, a noise function, and a simulated annotator bias.
Following prior work on simulated user preferences, we model the probability of a "correct" decision (the text with a higher latent score wins) as a function of the score difference and calibrate its parameters to match a target accuracy~\cite{simpson_interactive_2020, dubois_alpacafarm_2023, engelmann_arts_2024}.
The noise pattern follows a distance-calibrated curve $P(\text{correct}\mid \Delta)=\tfrac{1}{2}+(p_{\max}-\tfrac{1}{2})(1-e^{-\Delta/\tau})$, where $\tau$ is chosen so that $P$ matches a target accuracy at a reference difference.
For the simulation, we selected $p_{\max}$ = 0.99 to give a slight chance of an incorrect decision even with a large score delta. 
We set $\tau$ to 0.9, with the goal of achieving a win probability of 80\% at a score $\Delta$ of 90.
These settings produce a noise pattern that better reflects subjectivity scaling with the score delta than previous simulations with a fixed 10\% error rate\cite{dubois_alpacafarm_2023, engelmann_arts_2024}.
\Cref{fig:win_prob} shows a plot of the resulting win probabilities.
Aside from a distance-calibrated noise pattern that simulates the indecisiveness of annotators, we implement an annotator bias in our simulation.
This intuition is grounded in the premise that an annotator (human or model) has systematic biases that influence their perception of specific features as less or more biased relative to the latent scores.
Annotator bias is simulated by randomly choosing a number of $t_{bias}$ items, whose inherent scores are always treated as $\Delta_{bias}$ higher or lower than the latent severity in comparisons.
For each distribution configuration, we simulate $t_{bias}$ as 0, 50, and 200, for $\Delta_{bias} = 200$.
In combination, the distance-calibrated noise and these item-specific shifts approximate several annotator effects relevant to media bias annotation, including inconsistent or erratic decisions on close cases and systematic misjudgment.

\textbf{Evaluation Setup.}
We report cost as \emph{cost-equivalent API calls}, counting one call per pairwise decision or listwise permutation, not taking into account implied edges.
Due to the higher number of input tokens for listwise ranking calls, we calculate it as the number of listwise comparisons multiplied by half of the number of texts per list.
This way, we can use cost-equivalent API calls as a precise measure for calculation cost. 
Furthermore, it provides a reasonable estimate for human annotation effort.
In \Cref{sec:application}, we give a rough estimate of the cost reduction between different implementation variants in the application with real data.
Effectiveness is calculated based on the scores produced by the chosen rating system.
Here, Spearman’s $\rho$ is used to measure quality against the latent order in the simulation.

We use Elo for online scoring and matchmaking decisions for all approaches.
Elo is calculated under the assumption that the probability of text $T_1$ winning against text $T_2$ is determined by the difference in their Elo ratings.
If $T_1$  and $T_2$ have identical ratings, the estimated probability of $T_1$ winning is 50\%.
All Elo scores of our texts are initialized with the same predefined Elo value.
The expected probability $E_{T_1}$ that $T_1$ (Elo rating $R_1$) wins against $T_2$  (Elo rating $R_2$) is defined as \cite{e5b967c2-28fb-332e-93f5-e47fe8d1ccd0, boubdir2023elo}:

\begin{equation}
    E_{T_1} = \frac{1}{1 + 10^{(R_2-R_1)/400}}.
    \label{eq:1}
\end{equation}

After $T_1$ has won against $T_2$, the new rating $R^{\prime}_{1}$ of $T_1$ is:

\begin{equation}
    R^{\prime}_{1} = R_{1} + k (S_{T_1}- E_{T_1}) .
        \label{eq:2}
\end{equation}

The constant $k$ controls the degree of change after a game and is set to 32 for our simulation. 
$S_T$ is 1 if $T$ has won and 0 otherwise. 
For calculating the expected probability of $T_2$, we proceed in the same manner.
Initially, the ratings of all texts are set to 1500. 
For listwise comparisons, the scores are updated based on the inferred pairwise comparisons.

In addition to Elo, we apply the Bradley-Terry (BT) model for paired comparisons with a logistic link and maximum-likelihood estimation via the Minorization-Maximization updates to the sets of comparisons produced by all approach configurations~\cite{hunter_mm_2004}.
Items $i = 1, \ldots, n$ have positive “abilities” $\pi_i > 0$  (or equivalently scores $\theta_i = \log \pi_i$). 
For any pair $(i,j)$,
\begin{equation}
\begin{aligned}
     \Pr(i \succ j) & = \frac{\pi_i}{\pi_i+\pi_j}=\sigma(\theta_i-\theta_j),
\text{with} \\ 
    \qquad\sigma(x)  & =\frac{1}{1+e^{-x}}.
\end{aligned}
\end{equation}  
From the set of comparisons we aggregate wins $W_{ij}$ (the number of times $i$ beat $j$) and total matches $m_{ij}=W_{ij}+W_{ji}$.
The log-likelihood is
\small
\begin{equation}
    \ell(\pi) =\sum_{i<j} \big[\, W_{ij}\log \pi_i + W_{ji}\log \pi_j - m_{ij}\log(\pi_i+\pi_j) \,\big].
\end{equation}
\normalsize
As proposed by \citet{hunter_mm_2004}, we estimate $\pi$ by Minorization-Maximization iterations (monotone ascent of $\ell$):
\begin{equation}
    \pi_i^{(t+1)}  =  \frac{W_i + \lambda}{\displaystyle \sum_{j\neq i}\frac{m_{ij}}{\pi_i^{(t)}+\pi_j^{(t)}} + \lambda},
\qquad W_i=\sum_{j} W_{ij}.
\end{equation}
We set $\lambda$, a ridge-style stabilizer for sparse comparison graphs, to $10^{-6}$.
After each update, we normalize the geometric mean of $\pi$ to 1. 
BT scores are then calculated as $\theta_i=\log \pi_i - \tfrac{1}{n}\sum_k \log \pi_k$.

\begin{figure*}[t]
    \centering
    \begin{subfigure}[t]{0.47\textwidth} 
        \centering
        \includegraphics[width=\linewidth]{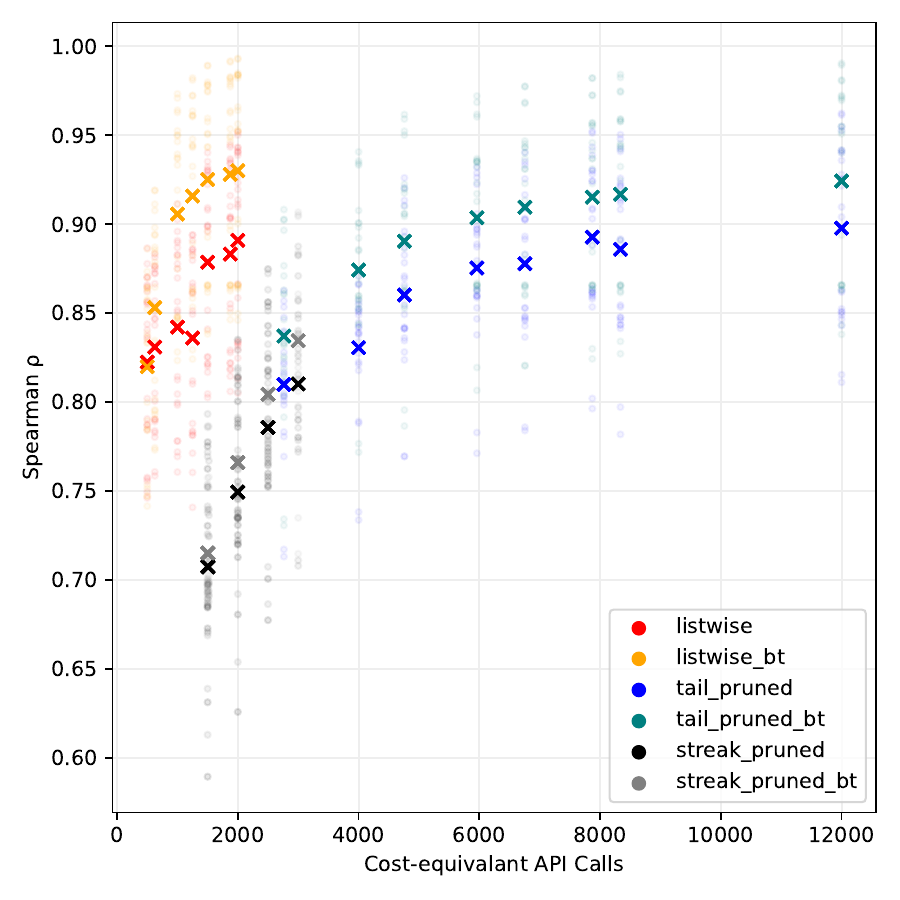}
        \caption{}
        \label{fig:simulation_results}
    \end{subfigure}
    \begin{subfigure}[t]{0.47\textwidth} 
        \includegraphics[width=\linewidth]{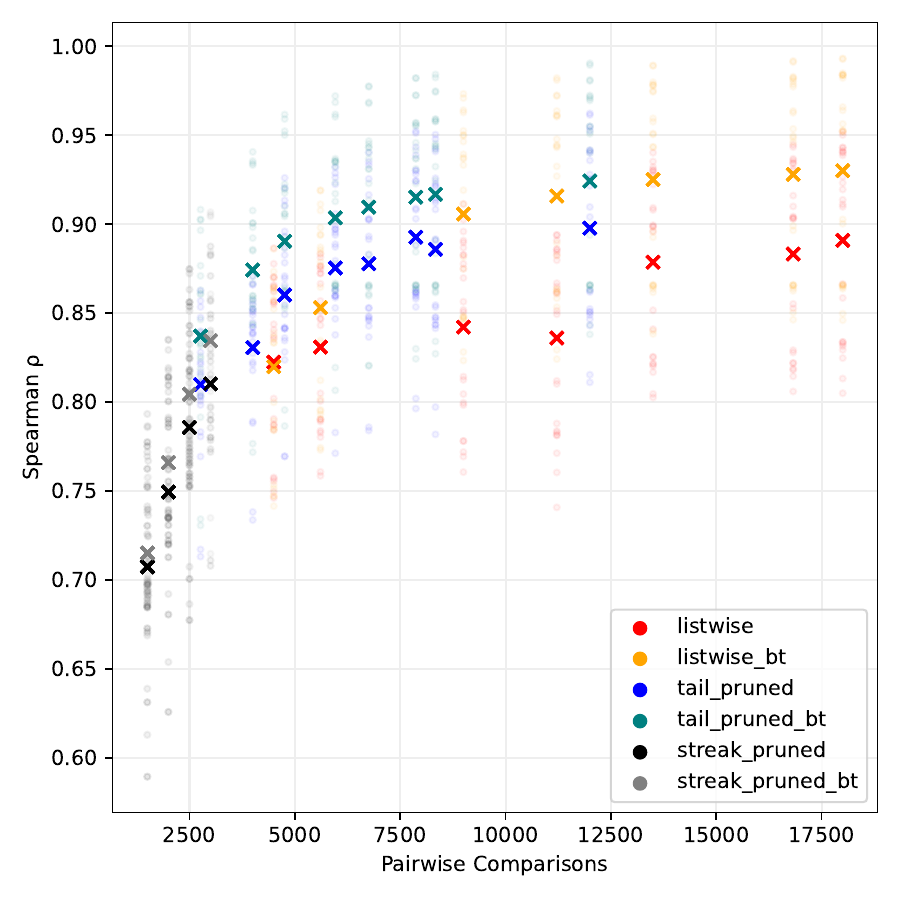}
        \caption{}
        \label{fig:simulation_results_pc}
    \end{subfigure}
    \begin{subfigure}[t]{1\textwidth}
    \centering
    \includegraphics[width=\linewidth]{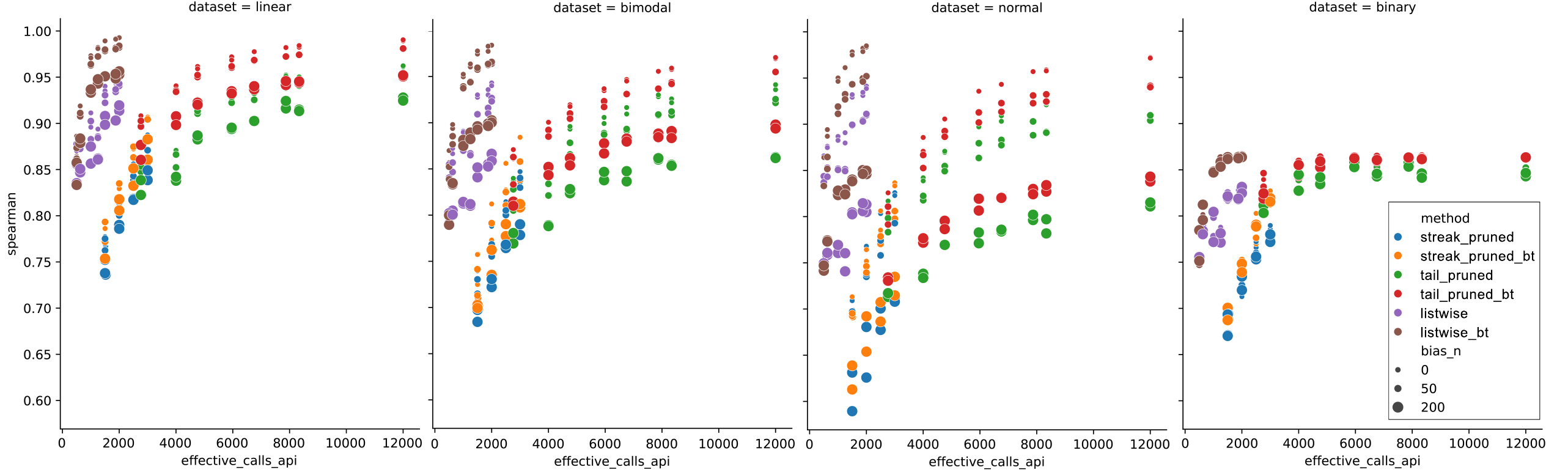}
    \caption{}
    \label{fig:simulation_results_by_distr}
\end{subfigure}
    \caption{Results of the simulation of different pairwise comparison implementations for a range of dataset configurations. (a) Spearman~$\rho$ rank correlation by cost as cost-equivalent API calls (means of each configuration marked by ``x''). (b) Spearman~$\rho$ rank correlation of simulations by number of pairwise comparisons, for listwise approaches inferred from list rankings (means of each configuration marked by ``x''). Pairwise comparison counts of listwise comparison approaches are inferred from listwise rankings. (c) Results of the simulation, separated by dataset configuration. The size of the data points indicates the strength of the simulated annotator bias.}
\end{figure*}

\textbf{Simulation Results. }
\Cref{fig:simulation_results} displays the simulation results, plotted as rank correlation and cost-equivalent API calls on the x-axis.
\Cref{fig:simulation_results_pc} shows the same correlation scores, sorted by the number of pairwise comparisons. 
\Cref{fig:simulation_results_by_distr} visualizes the scores separated by value distribution in the dataset, with the different model bias simulations shown in the plots.
Note that in all cases, the rightmost tail-pruning configuration is unpruned and thus represents a baseline of 24 rounds of pairwise comparisons without modification.

Overall, the more pairwise comparisons (inferred or directly assessed), the better the rank correlation.
In all approaches and for every simulated dataset setup, BT scoring outperforms Elo-based scoring, and matchmaking predicated on score similarity consistently surpasses random matchmaking.
\Cref{fig:simulation_results_pc} reveals that while inferred pairwise comparisons from listwise comparisons lead to overall worse results for the same number of comparisons, \Cref{fig:simulation_results} shows that cost-wise, using listwise comparison approaches appears to be much more efficient.
Listwise comparisons benefit most from using BT rating, with the highest average difference between mean scores of BT and Elo scores for each configuration.
As shown in \Cref{fig:simulation_results_by_distr}, the distribution of latent scores primarily affects the convergence point of Spearman~$\rho$ across methods and the variance observed across method configurations and simulated biases.
Simulated annotator bias has the strongest effect when the latent scores are normally distributed.

To compare alternatives, we normalize both cost $c$ (cost-equivalent API calls) and effectiveness (Spearman~$\rho$) $e$, and calculate $score_\alpha = \alpha \cdot e + (1-\alpha) \cdot (1-c)$ for $\alpha = 0.4$, putting slightly more weight on effectiveness.
Using this scoring, we find three rounds of listwise comparisons with 10 elements and no overlap between lists to be the best alternative (score: 0.95, cost-equivalent API calls: 1,500, Spearman~$\rho$: 0.93), outperforming the baseline approach of 24 rounds of pairwise comparison (cost-equivalent API calls: 12,000, Spearman~$\rho$: 0.92).
Six more listwise approaches then follow this configuration.
The best non-listwise alternative is tail pruning with 24 rounds of comparisons, pruning starting after 8 rounds, and pruning of 20\% (score: 0.72, cost-equivalent API calls: 4,759, Spearman~$\rho$: 0.89).

\section{Application of Pairwise Comparison}\label{sec:application}
Simulation can help identify promising candidates for cost-efficient alternatives.
The results allow us to compare the effectiveness of these approaches under controlled (simulated) conditions.
However, we cannot directly deduce their suitability for bias detection and quantification.
To evaluate the pairwise comparison and cost-aware variants for these tasks, we apply the method to two benchmarking datasets~\citet{Wessel2023}.\footnote{All data, prompts, and code required to reproduce our results can be found at \url{https://github.com/HaakFab/bias_pc}.} 

All comparisons are carried out by LLMs using short prompts that \begin{inparaenum}[i)]
\item define the target concept of bias for the respective dataset, 
\item describe the comparison task (pairwise choice or listwise ranking), and 
\item constrain the output format (see repository for full templates).
\end{inparaenum} 
In this work, we use a single, carefully designed prompt per task and model, and we do not systematically vary prompt wording, since we focus on the general applicability of pairwise comparison.
A more extensive study of prompt sensitivity is left for future work.

\subsection{Bias Detection}\label{sec:detection}
The advantages of pairwise comparison-based rating are primarily used for quantification.
Nevertheless, by defining a threshold score, we can interpret the resulting ratings as binary labeling decisions.
By testing how well pairwise comparison can differentiate between biased and unbiased texts, we can determine its suitability for this purpose.
For evaluation, we use the BABE dataset, a high-quality dataset comprising 3,673 sentences from news texts labeled by trained expert annotators~\cite{spinde2021babe}.

We use the most promising configuration from the simulation, which involves three rounds of listwise ranking with lists of 10 texts, utilizing both \emph{GPT-5-nano} and \emph{GPT-5-mini}\footnote{Non-structural prompt (full prompts can be found in the repository): \emph{"You are an expert in analyzing bias. You will receive K short texts, each with a numeric id 0..K-1.Rank them from MOST biased to LEAST biased."}}.
For comparison, we produced a run of 24 rounds of pairwise comparison with \emph{GPT-5-nano}.
Since this approach is much more comparison-intensive than the listwise approach and we can directly assess the difference between a large amount of listwise comparisons and the much more efficient listwise comparison approach with the \emph{GPT-5-nano} model, we decided not to use the larger mini model for cost reasons. 
As a primary baseline, we use bias assessments by \emph{GPT-5-mini}.
Moreover, our findings are evaluated against the best-performing method detailed by \citet{spinde2021babe} in the BABE publication, which employs a $BERT$ classifier with distant supervision.

\Cref{tab:detection} shows the results of the evaluation.
Binary labels from Elo scores are interpreted as biased if the score is larger than the starting Elo.
Since an Elo score larger than the starting Elo is a result of a text being generally labeled as more biased more frequently than less biased, we assume the starting Elo to be a good pivot point for a binary decision.
For BT, the score is considered biased if the $\theta$ rating is positive ($\theta$ always has a mean of zero).
The results show that the BT-produced scores of the listwise-comparison \emph{GPT-5-mini} approach outperform all baselines, even though the scores are very similar.
The difference between the listwise and pairwise comparison approaches is minimal, which we interpret as confirmation of our simulation results.
The baseline of direct assessment by \emph{GPT-5-mini} is not far off from our best-performing approach, but it lacked the additional relational information contained in the ratings from which the binary label was interpreted.
To evaluate this capability, we apply our approach to a dataset with labels on a continuous scale in \Cref{sec:quantification}.


\begin{table*}[t]
\centering
\begin{tabular}{l l l l l}
\toprule
                                                             Approach      & Recall         & Accuracy       & Precision      & Macro \emph{F1}             \\ 
\midrule
$mini\_direct$                                                        & \textbf{0.851} & 0.780           & 0.741          & 0.792          \\
$BERT + distant$ &&&& 0.804 \\
\midrule
$nano\_24\_pairwise_{Elo}$                                                      & 0.792          & 0.790           & 0.784          & 0.788          \\
$nano\_24\_pairwise_{BT}$ & 0.776          & 0.796          & \textbf{0.803}          & 0.790           \\
$mini\_listwise_{Elo}$                                                       & 0.783          & 0.779         & 0.772          & 0.778          \\
$mini\_listwise_{BT}$                                                    & 0.844          & \textbf{0.803}          & 0.776          & \textbf{0.808} \\
$nano\_listwise_{Elo}$                                                       & 0.761          & 0.763          & 0.756          & 0.761          \\
$nano\_listwise_{BT}$                                                    & 0.796          & 0.784          & 0.773          & 0.784 \\
\bottomrule
\end{tabular}
\caption{Recall, Accuracy, Precision, and Macro F1 results of predicted labels for the BABE dataset, produced by 24 rounds of pairwise comparisons and the cost-aware alternative of 3 rounds of listwise comparison, as well as two baseline approaches (direct assessment by LLM, and $BERT + distant$ by \citet{spinde2021babe}).}
\label{tab:detection}
\end{table*}

\subsection{Bias Quantification}
\label{sec:quantification}
The primary advantage of pairwise comparison-based rating approaches is that the resulting scores are a direct quantification of a linguistic feature.
Through various straightforward comparative evaluations, we can rate objects (or, in this instance, texts) based on the strength of a particular characteristic.

To assess the efficacy of pairwise comparison, specifically the chosen cost-aware variants, we use them to label the test set from the ``Us vs. Them'' dataset, as described by \citet{huguet_cabot_us_2021}.
The dataset is constructed from Reddit posts that have been annotated through crowd-sourcing methods.
Texts are annotated as \emph{supportive}, \emph{neutral}, \emph{critical}, and \emph{discriminatory}.
Labels are then aggregated as the mean of all annotators.
Based on the aggregated individual labels, all texts are assigned a continuous scale from 0~(\emph{supportive}) to 1~(\emph{discriminatory}), with $1/3$~meaning \emph{neutral} and $2/3$~\emph{critical}~\cite{huguet_cabot_us_2021}. 
We strongly question the quality of continuous labels produced in this manner of deriving continuous scores from a vaguely connected nominal scale.
However, due to the lack of alternatives, we decided to use the dataset for evaluating the quantification capabilities of our approach.
As for the detection evaluation in \Cref{sec:detection}, we compared our approach with a baseline of direct assessment in the form of a score assigned by \emph{GPT-5-mini}.
We did not produce pairwise comparison ratings because we determined that the listwise cost-aware approach had been adequately proven to produce similar results through simulation and application with the BABE corpus.
Due to the nature of the scale, we were compelled to more carefully define ``bias'' in our prompts, both for the baseline and our approaches.\footnote{ Bias definition for "Us vs. Them" dataset: \emph{``You are an expert judge of bias in the sense of negative sentiment and stance towards minorities or demographic groups. Given a list of short texts, rank them from MOST biased (Discriminatory) to LEAST biased (Supportive).''}}

\Cref{tab:quantification} shows the results as the Pearson~$r$ correlation between the produced scores and the dataset's labels.
The direct assessment baseline outperforms the listwise approaches, although only by a small margin in the case of the BT-rated \emph{GPT-5-mini} labels.
However, if we only look at the texts labeled ``critical'' or ``discriminatory'', a label based on the mean of all scores and intended for binary classification, \emph{GPT-5-mini} with BT rating outperforms the baseline. 
\emph{GPT-5-nano} performed notably worse than \emph{GPT-5-mini} in listwise approaches.
These equally moderate Pearson~$r$ correlations for pairwise comparison and direct assessment are likely influenced by the way the ``Us vs. Them'' labels are constructed.
This suggests that part of the gap may be due to limitations of the targets themselves rather than purely to shortcomings of the scoring methods.

\begin{table}[t]
\normalsize
\begin{tabular}{l l l}
\toprule
                                 Approach      & Full dataset   & Critical texts \\ 
\midrule
$mini\_direct$                               & \textbf{0.542} & 0.505           \\
\midrule
$mini\_listwise_{Elo}$ & 0.510          & 0.500            \\
$mini\_listwise_{BT}$     & 0.537          & \textbf{0.536}  \\
$nano\_listwise_{Elo}$ & 0.349          & 0.431          \\
$nano\_listwise_{BT}$     & 0.346          & 0.460          \\
\bottomrule
\end{tabular}
\caption{Pearson $r$ correlations of ratings produced by our approaches and the direct assessment baseline with the labels from the ``Us vs. Them'' dataset. }
\label{tab:quantification}
\end{table}

\subsection{Evaluation}

While the results of bias detection using pairwise comparison (\Cref{sec:detection}) demonstrate the method's capabilities, the main advantage of pairwise comparison can only be superficially evaluated using the available datasets. 
For quantification (\Cref{sec:quantification}), absolute correlations are modest across all methods, likely reflecting label construction issues in that corpus (discrete stance categories aggregated into a pseudo-continuous score). 
Even so, listwise comparison with BT rating utilizing \emph{GPT-5-mini} tracks the direct-assessment baseline closely overall and exceeds it in the segment of critical or discriminatory texts (\Cref{tab:quantification}). 
This mirrors the simulation findings: similarity-based matchmaking and BT estimation yield stable orderings under tight budgets, and listwise grouping converts local rankings into many implied edges, delivering strong cost-quality trade-offs. 

Direct assessment with current LLMs is competitive in terms of raw accuracy and cost-effectiveness.
However, it is possible that the datasets are part of the training data for the models and the scores are recalled rather than produced.
Furthermore, pairwise comparison-based quantification approaches have the practical advantage of being more transparent and explainable.
Concrete rankings and inferred comparisons back every score, enabling audit trails, targeted error analysis, and thresholding policies. 
We rely on two models from the same family rather than on a broad model sweep.
While this is sufficient for our primary goal of establishing the feasibility of cost-aware pairwise comparison for bias detection and quantification, the consistent gains of \textit{GPT-5-mini} over \textit{GPT-5-nano} across settings indicate that model choice does matter and that stronger models can yield higher-quality comparison signals.
We did not explore the effects of different prompts and bias definitions. 
Tuning both for a specific task or dataset could, in theory, improve the effectiveness of pairwise comparison.
This is especially true for bias, which can take a wide range of forms~\cite{spinde_media_2024}. 
Detailed exploration of model diversity and prompt setup is left for future work.





    
    




\section{Conclusion and Outlook}
In this work, we addressed the methodological and practical challenges of bias assessment by deriving continuous bias judgments from sequences of pairwise comparisons aggregated using probabilistic rating procedures. 
We explored the effects of various rating techniques and the parameters of cost-aware alternatives under controlled (simulated) conditions. 
Our results suggest that similarity-based matchmaking systematically improves sample efficiency over random matchmaking by concentrating comparisons where uncertainty is highest and information gain is largest. 
Listwise grouping converts local rankings into many implied edges, delivering strong cost-quality trade-offs. 
Our findings support the use of the method as a practical foundation for quantifying subjective linguistic aspects, enabling reproducible bias analysis.

Offline Bradley-Terry estimation offers stronger rank stability than online Elo rating under identical comparison budgets.
While streak and tail pruning provide reasonable alternatives, listwise comparison stands out as the most promising cost-aware approach.
By converting rankings of ten items into many implied pairwise edges, it substantially reduces labeling costs with no significant loss in label quality. 
Therefore, it is a good default when input tokens or human annotation budgets are the primary concerns.  

To assess whether the results of our simulation hold up when applied to real data and to evaluate the applicability of pairwise comparisons for bias detection and quantification, we tested our approach against two human-annotated datasets with the following results: 
\begin{inparaenum}[A)]
    \item In detection (\Cref{sec:detection}), the most promising cost-aware configuration, three listwise ranking rounds with 10-item groups, matches or surpasses the effectiveness of generic pairwise comparison while reducing cost by an order of magnitude.
    In particular, listwise comparison with BT rating \emph{GPT-5-mini} achieves the best overall balance and slightly outperforms both baselines, direct-assessment LLM and BERT+distant-supervision (\Cref{tab:detection}). 
    \item For quantification (\Cref{sec:quantification}), absolute correlations are modest across all methods, likely reflecting label construction issues in that corpus (discrete stance categories aggregated into a pseudo-continuous score). 
\end{inparaenum}
Even so, listwise comparison by GPT-5-mini judgments with BT rating is on par with the direct-assessment baseline and exceeds it on the practically salient segment of critical/discriminatory texts (\Cref{tab:quantification}).

Future work could enhance the capabilities of binary classification through pairwise comparison and threshold optimization, particularly for unweighted label distributions.
However, binary detection does not use this technique to its fullest potential.
This is true in particular because more capable LLMs can solve this task more efficiently.
The effects of prompts on labeling decisions, the inclusion of more diverse model selections, and the use of personas are open to investigation in future research.
Multi-persona or model ensembles, or rotating annotator pools, could be integrated into the same pairwise comparison framework to mitigate idiosyncratic biases while maintaining the sample-efficient nature of the approach.
Finally, the (optimized) pairwise comparison approach could be used to facilitate the creation of high-quality, human-annotated datasets with continuous bias severity labels. 

Overall, optimizing pairwise comparison provides a scalable and principled path to identify and quantify bias, delivering cost-aware pipelines that retain interpretability and are compatible with both human and LLM annotators.

\appendix

\bibliography{bibliography}

\end{document}